\documentclass[sigconf]{acmart}

\usepackage{amsmath,amsfonts}  
\usepackage{algorithmic}
\usepackage{graphicx}
\usepackage{textcomp}
\usepackage{xcolor}
\usepackage{enumitem}
\usepackage{mathtools}
\usepackage[ruled,vlined]{algorithm2e}
\usepackage{booktabs}
\usepackage[normalem]{ulem}
\usepackage{graphicx}
\usepackage{subcaption}
\usepackage{marvosym}
\usepackage{balance}
\usepackage{hyperref}
\usepackage{pifont}
\usepackage{multirow}
\usepackage[skip=1pt]{caption}

\usepackage{xcolor}
\usepackage{pifont}

\definecolor{darkgreen}{RGB}{50,100,0}
\definecolor{darkred}{RGB}{200, 0, 0}
\newcommand{\CheckmarkBold}{\textcolor{darkgreen}{\ding{51}}} %
\newcommand{\XSolidBrush}{\textcolor{darkred}{\ding{55}}} 

\newtheorem{definition}{Definition}
\newtheorem{theorem}{Theorem}
\newtheorem{problem}{Problem}
\usepackage{balance}

\AtBeginDocument{%
  }

\copyrightyear{2024}
\acmYear{2024}
\setcopyright{rightsretained}
\acmConference[CIKM '24]{Proceedings of the 33rd ACM International
Conference on Information and Knowledge Management}{October 21--25,
2024}{Boise, ID, USA}
\acmBooktitle{Proceedings of the 33rd ACM International Conference on
Information and Knowledge Management (CIKM '24), October 21--25, 2024,
Boise, ID, USA}
\acmDOI{10.1145/3627673.3679696}
\acmISBN{979-8-4007-0436-9/24/10}

\begin{document}

\title{FCS-HGNN: Flexible Multi-type Community Search in Heterogeneous Information Networks}

\author{Guoxin Chen}
\affiliation{%
  \institution{Key Laboratory of AI Safety, Institute of Computing Technology, CAS}
  \institution{University of Chinese Academy of Sciences}
  \city{BeiJing}
  \country{China}}
\email{chenguoxin22s@ict.ac.cn}

\author{Fangda Guo}
\authornote{Corresponding author}
\affiliation{%
  \institution{Key Laboratory of AI Safety, Institute of Computing Technology, CAS}
  \city{BeiJing}
  \country{China}}
\email{guofangda@ict.ac.cn}

\author{Yongqing Wang}
\affiliation{%
  \institution{Key Laboratory of AI Safety, Institute of Computing Technology, CAS}
  \city{BeiJing}
  \country{China}}
\email{wangyongqing@ict.ac.cn}

\author{Yanghao Liu}
\affiliation{%
  \institution{Key Laboratory of AI Safety, Institute of Computing Technology, CAS}
  \institution{University of Chinese Academy of Sciences}
  \city{Beijing}
  \country{China}}
\email{liuyanghao19s@ict.ac.cn}

\author{Peiying Yu}
\affiliation{%
  \institution{Natural Language Processing Lab, School of Computer Science \& Technology, Soochow University}
  \city{Suzhou}
  \country{China}}
\email{peiying.yu.chn@gmail.com}

\author{Huawei Shen}
\author{Xueqi Cheng}
\affiliation{%
  \institution{Key Laboratory of AI Safety, Institute of Computing Technology, CAS}
  \city{BeiJing}
  \country{China}}
\email{{shenhuawei,cxq}@ict.ac.cn}


\renewcommand{\shortauthors}{Guoxin Chen et al.}

\begin{abstract}
Community search is a personalized community discovery problem designed to identify densely connected subgraphs containing the query node.
Recently, community search in heterogeneous information networks (HINs) has received considerable attention.
Existing methods typically focus on modeling relationships in HINs through predefined meta-paths or user-specified relational constraints.
However, metapath-based methods are primarily designed to identify single-type communities with nodes of the same type rather than multi-type communities involving nodes of different types.
Constraint-based methods require users to have a good understanding of community patterns to define a suitable set of relational constraints, which increases the burden on users.
In this paper, we propose FCS-HGNN, a novel method for flexibly identifying both single-type and multi-type communities in HINs.
Specifically, FCS-HGNN extracts complementary information from different views and dynamically considers the contribution of each relation instead of treating them equally, thereby capturing more fine-grained heterogeneous information.
Furthermore, to improve efficiency on large-scale graphs, we further propose LS-FCS-HGNN, which incorporates i) the neighbor sampling strategy to improve training efficiency, and ii) the depth-based heuristic search strategy to improve query efficiency.
We conducted extensive experiments to demonstrate the superiority of our proposed methods over state-of-the-art methods, achieving average improvements of 14.3\% and 11.1\% on single-type and multi-type communities, respectively.\footnote{The code is available at \url{https://github.com/Chen-GX/FCS-HGNN}}
\end{abstract}




\maketitle

\section{Introduction}
\label{section:introduction}
Community search, a kind of query-dependent personalized community discovery problem designed to identify densely-connected subgraphs containing query nodes, has garnered considerable interest~\cite{luo2020efficient,guo2021multi,chen2023communityaf,DBLP:conf/sigir/CaiZLW23,chen2022vics,miao2022reliable,wang2024attribute,chen2023ics}.
Most existing studies primarily focus on homogeneous graphs associated with a single type of nodes and edges.
However, many real-world data are naturally represented as heterogeneous information networks (HINs), where multiple types of entities and the relations between them are embodied by various types of nodes and edges, respectively.
Compared with homogeneous graphs, HINs possess richer semantics and complex relationships, encapsulating more valuable community information.

\begin{figure}[t]
    \centering
    \includegraphics[width=\linewidth]{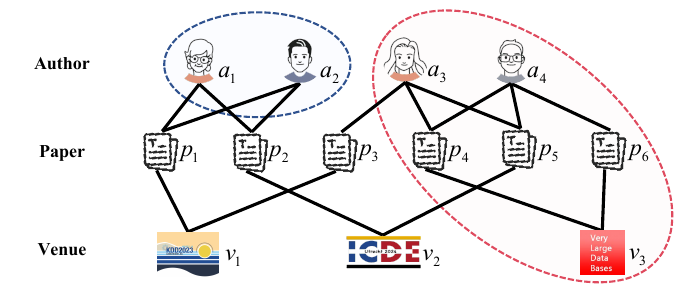}
    \caption{The \textcolor{blue}{blue} and \textcolor{red}{red} circles represent single-type and multi-type community examples in the bibliographic network, respectively.}
    \label{fig:example_of_hin}
    \vspace{-10pt}
\end{figure}

Recent advances in community search over HINs usually endeavor to model the relationships between nodes of different types through the predefined meta-paths~\cite{fang2020effective,qiao2021keyword,jiang2022effective,zhou2023influential,liu2023significant} or user-specified relational constraints~\cite{jian2020effective}.
However, as shown in Table~\ref{tab:difference}, they suffer from four main limitations.
\textbf{(1)} These metapath-based methods are primarily designed to identify single-type communities with nodes of the same type, rather than multi-type communities involving nodes of different types~\cite{jian2020effective}.
The single-type community does not fully exploit the valuable community information in HINs.
For example, as shown in Figure~\ref{fig:example_of_hin}, the single-type community composed of authors (blue circle) can only uncover collaboration information among authors.
Instead, a multi-type community spanning authors, papers, and venues (red circle) may reflect the core of a research field. By analyzing such multi-type communities, we can not only understand collaboration patterns among authors but also delve deeper into research trends by papers and influential venues in this field, thereby more effectively supporting various downstream applications~\cite{chakraborty2015formation,li2018team,gajewar2012multi,behrouz2022firmtruss,chen2023mprompt,chen2024seer,chen2024alphamath}.
\textbf{(2)} Constraint-based methods require users to have a good understanding of community patterns, in order to define a suitable set of relational constraints. Otherwise, without any guidance, they may fail to find any communities, which significantly increases the burden on end-users.
\textbf{(3)} Both metapath-based and constraint-based methods suffer from the pattern inflexibility, which refers to the problem that they are based on predefined community patterns, such as $(k, \mathcal{P})$-core~\cite{fang2020effective,zhou2023influential,liu2023significant}, and $(k, \Psi)$-NMC~\cite{jiang2022effective}, or the user-specified relational constraints.
In real-world scenarios, the pattern of a community is flexible in nature, and the target communities may not conform to predefined community patterns or relational constraints, thus risking being overlooked.
\textbf{(4)} Both metapath-based and constraint-based methods only focus on structural properties of HINs~\cite{fang2020effective,jian2020effective,jiang2022effective} or limited low-dimensional node features~\cite{zhou2023influential,liu2023significant}.
For example, \cite{zhou2023influential} only considers one-dimensional features (such as influence).
In practical scenarios, HINs usually naturally endow nodes with high-dimensional features containing important information, which is a challenge for existing methods in HINs.

\begin{table}[t]
\centering
\caption{Comparison with other CS Methods in HINs.}
    \resizebox{\linewidth}{!}{
\begin{tabular}{@{}cccccc@{}}
\toprule
\textbf{Method}                                     & \textbf{\begin{tabular}[c]{@{}c@{}}Single\\ type\end{tabular}} & \textbf{\begin{tabular}[c]{@{}c@{}}Multi\\ type\end{tabular}} & \textbf{\begin{tabular}[c]{@{}c@{}}Node\\ feature\end{tabular}} & \textbf{Meta-path} & \textbf{\begin{tabular}[c]{@{}c@{}}Community\\ patterns\end{tabular}} \\ \midrule
e.g.,~\cite{fang2020effective,jiang2022effective}   & \CheckmarkBold                                                 & \XSolidBrush                                                  & \XSolidBrush                                                    & Predefined         & Predefined                                                            \\
e.g.,~\cite{zhou2023influential,liu2023significant} & \CheckmarkBold                                                 & \XSolidBrush                                                  & \CheckmarkBold(Low Dim.)                                        & Predefined         & Predefined                                                            \\
e.g.,~\cite{jian2020effective}                      & \XSolidBrush                                                   & \CheckmarkBold                                                & \XSolidBrush                                                    & Not Dependent      & User-specified                                                        \\ \midrule
\textbf{Ours}                                       & \CheckmarkBold                                                 & \CheckmarkBold                                                & \CheckmarkBold                                                  & Not Dependent      & Data driven                                                           \\ \bottomrule
\end{tabular}
}
\label{tab:difference}
\end{table}

To address the above limitations, we propose FCS-HGNN, a query-driven heterogeneous graph neural network that can flexibly identify multi-type communities, as well as single-type communities in HINs.
To tackle pattern inflexibility, FCS-HGNN adaptively learns community patterns in a data-driven manner rather than relying on strict predefined community patterns or user-specified relational constraints.
Specifically, we design two encoders: (1) \textit{Heterogeneous Encoder} to deal with the heterogeneity and learn query-independent node features, and (2) \textit{Query Encoder} to learn the local structural information around the query nodes.
In each encoder, we introduce an edge semantic attention mechanism to dynamically consider the contribution of each relation when identifying diverse multi-type communities rather than treating them equally.
It is worth noting that graph neural networks offer a unified way to combine the structural information and the arbitrary dimensional node features.
To take advantage of complementary information from different encoders, FCS-HGNN integrates them through an attention mechanism after each layer and ultimately utilizes them in the final layer to infer the probability of each node belonging to the community.
Finally, by considering the nodes' probabilities, we employ a breadth-first search (BFS) algorithm to obtain the target multi-type community.

Furthermore, to improve efficiency on large-scale graphs, we introduce the neighbor sampling algorithm and depth-based heuristic search strategy to further optimize our algorithm, named LS-FCS-HGNN (\textbf{L}arge \textbf{S}cale FCS-HGNN).
Through the neighbor sampling algorithm, in each iteration, LS-FCS-HGNN updates its parameters on a randomly sampling subgraph, rather than on the entire large-scale graph, which significantly reduces the computational overhead and improves training efficiency.
Furthermore, based on real-life observations,  we find that the nodes within the same community typically exhibit more edges and shorter paths. In other words, the nodes closer to the query node are more likely to belong to the community.
Therefore, we propose a depth-based heuristic search strategy to reduce searches for the nodes farther from the query node, thereby significantly improving query efficiency.

To summarize, we make the following contributions.

\noindent $\bullet$ We propose FCS-HGNN, a novel query-driven heterogeneous graph neural network capable of effectively identifying multi-type communities, as well as single-type communities in heterogeneous information networks.
FCS-HGNN adaptively learns community patterns in a data-driven manner, without relying on predefined community patterns and user-specified relational constraints, significantly alleviating the burden on end-users.

\noindent $\bullet$ We further propose LS-FCS-HGNN, an advanced algorithm designed to improve the efficiency in large-scale graphs.
LS-FCS-HGNN incorporates two strategies including i) the neighbor sampling algorithm to reduce computational overhead and improve training efficiency, and ii) depth-based heuristic strategy to improve the query efficiency, while preserving community effectiveness.

\noindent $\bullet$ We conduct extensive experiments on five real-world heterogeneous information networks to demonstrate the superiority of our proposed methods over state-of-the-art methods.
Our proposed LS-FCS-HGNN not only significantly enhance the effectiveness and flexibility of multi-typed community search but also obviously improve both training and query efficiency.


\vspace{-10pt}

\section{RELATED WORK}\label{section:related_work}


\textbf{Community Search in HINs.}
Recently, an increasing number of researchers have shown interest in community search over HINs~\cite{fang2021cohesive,fang2020effective,jian2020effective,qiao2021keyword,jiang2022effective,zhou2023influential,liu2023significant,chen2023causality}, due to the prevalence and significance of HINs in real-world scenarios.
These studies usually model the relationships between nodes in HINs through predefined meta-paths or user-specified relational constraints.
For example, \cite{fang2020effective} proposes to employ predefined meta-paths to delineate the connectivity among nodes of different types in HINs, and extend the $k$-core to $(k,\mathcal{P})$-core to measure the cohesiveness of a community, where $\mathcal{P}$ is the predefined meta-path.
Metapath-based methods aim to find single-type communities consisting of nodes with the same type, rather than discovering multi-type communities involving nodes of different types.
\cite{jian2020effective} proposes a relational community model, which defines communities based on a set of user-specified relational constraints.
It requires users to have a good understanding of community patterns to define a suitable set of constraints, which significantly increases the burden on users.
If the constraints are not specified correctly, it may fail to find any communities.
In contrast, our FCS-HGNN alleviates pattern inflexibility by adaptively learning community patterns in a data-driven manner, thereby alleviating the burden on end-users.

\textbf{Heterogeneous Graph Representation Learning.}
A large amount of work~\cite{wang2022survey} has been devoted to address the problem of learning node representations from HINs, including heterogeneous graph embedding~\cite{dong2017metapath2vec,fu2017hin2vec} and heterogeneous graph neural networks~\cite{wang2019heterogeneous,zhang2019heterogeneous,lv2021we,yang2023simple}.
Heterogeneous graph embedding primarily focuses on preserving structure information based on meta-paths.
For example, metapath2vec~\cite{dong2017metapath2vec} devised a metapath-based random walk and utilized skip-gram~\cite{mikolov2013efficient} to learn node representations.
Heterogeneous graph neural networks can be divided into metapath-based methods and metapath-free methods.
The design and selection of meta-paths still pose challenges in practice, which may require sufficient expert knowledge.
In contrast, our proposed FCS-HGNN adopts a metapath-free manner (\textit{Heterogeneous Encoder}) to deal with heterogeneity and employs the \textit{Query Encoder} to learn local structural information around query nodes, which sufficiently facilitates the learning of community information in HINs.


\section{Preliminaries}\label{section:preliminary}

\begin{definition}[Heterogeneous Information Network, HIN]
A HIN is a directed graph $\mathcal{H}=(V,E,\psi,\phi)$ with a node type mapping function $\psi: V \rightarrow \mathcal{A}$ and an edge type mapping function $\phi: E \rightarrow \mathcal{R}$, where $V$ and $E$ represent the sets of nodes and edges respectively, each node $v \in V$ belongs to a node type $\psi(v)\in \mathcal{A}$, each edge $e\in E$ belongs to an edge type $\phi(e)\in \mathcal{R}$, and $\vert\mathcal{A}\vert + \vert \mathcal{R} \vert > 2$.
\end{definition}
\vspace{-5pt}


\begin{definition}[Multi-type Community]\label{definition:multi-type community}
Given a HIN $\mathcal{H}=(V,E,\psi,\phi)$ and the set of node types $\mathcal{S}_{\mathcal{A}} \subseteq \mathcal{A}$, a multi-type community is defined as a subgraph $\mathcal{C}_{q} = (V_{q}, E_{q}, \psi, \phi)$ of $\mathcal{H}$ that satisfies the following criteria:
(1) Each node $v \in V_{q}$ satisfies $\psi(v) \in \mathcal{S}_{\mathcal{A}}$;
(2) $\mathcal{C}_q$ is connected and has a cohesive structure.
\end{definition}
\vspace{-5pt}

\begin{problem}[Multi-type Community Search in HINs]
    Given a HIN $\mathcal{H}=(V,E,\psi,\phi)$, a query node $v_q \in V$, and the set of node types $\mathcal{S}_{\mathcal{A}}$, multi-type community search problem aims to identify the query-dependent community $\mathcal{C}_{q} = (V_{q}, E_{q}, \psi, \phi)$ satisfying \textit{Multi-type Community} (refer to \textit{Definition~\ref{definition:multi-type community}}) and $v_q\in V_q$.
\end{problem}

Based on the above problem definition, we formalized the ML-based training manner as follows:
given the set of node types $\mathcal{S}_{\mathcal{A}}$ and a set of query tasks $\mathcal{D}=\{(q_1, pos(q_1), neg(q_1)), (q_2, pos(q_2),$ $ neg(q_2)), \cdots\}$, where $q$ represents the query node, $pos(q)$ and $neg(q)$ respectively represent the subsets of nodes that are in or not in the target community.
Our training objective is to learn a function $\mathcal{F}(\mathcal{H}, q, \mathcal{S}_{\mathcal{A}})$ that yields the probability of each node belonging to the target community.
Then, we employ search algorithm to find the target community with the highest probability.

\section{FCS-HGNN}\label{section:fcs-hgnn}
\begin{figure*}[ht]
    \centering
    \includegraphics[width=\linewidth]{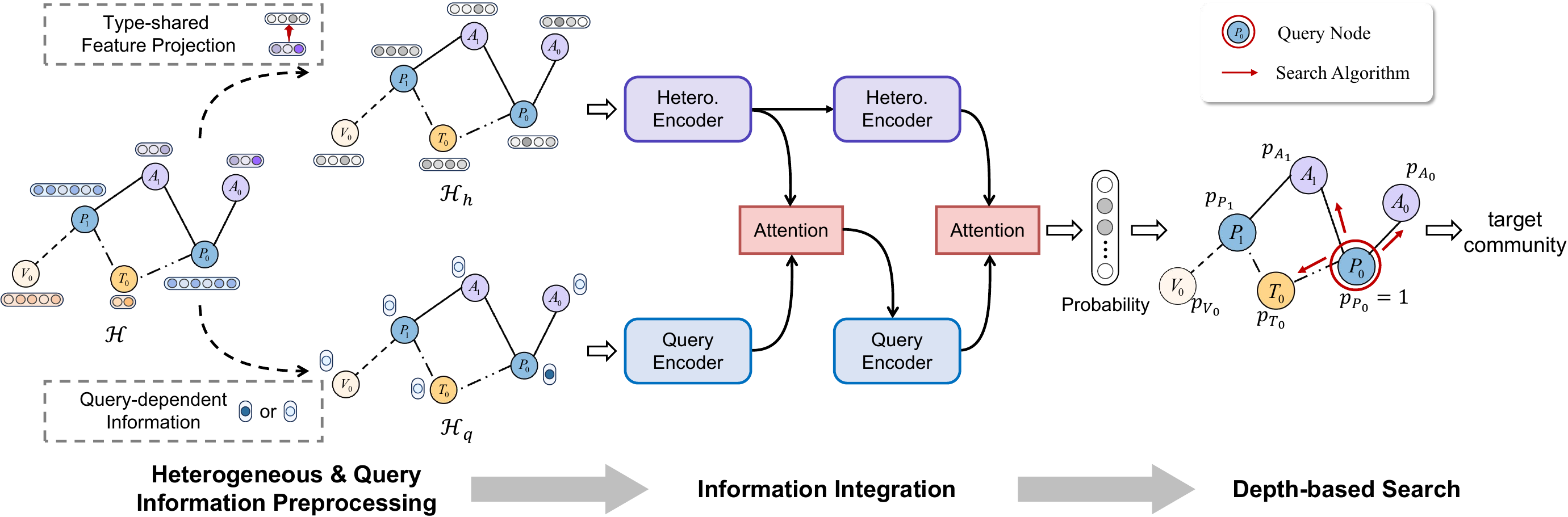}
    \caption{The overall framework of FCS-HGNN.}
    \label{fig:framework}
    \vspace{-1em}
\end{figure*}

We illustrate the overall framework of FCS-HGNN in
Figure~\ref{fig:framework}, which includes three main steps: \textit{Heterogeneous and Query Information Preprocessing}, \textit{Information Integration}, and \textit{Search Algorithm}.

\vspace{-5pt}

\subsection{Heterogeneous \& Query Information Preprocessing}
In the community search task over HINs, we primarily have to deal with two aspects of information: the query-dependent information provided by the user and the query-independent information inherent in HINs.
Differing from homogeneous graphs, HINs typically comprise multiple types of nodes and edges, presenting significant challenges due to their heterogeneity.
Existing works primarily address heterogeneity through predefined meta-paths~\cite{fang2020effective,qiao2021keyword,jiang2022effective,zhou2023influential,liu2023significant} or user-specified relationship constraints~\cite{jian2020effective}, which results in inflexibility and elevates the burden on end-users.

\subsubsection{Type-shared Feature Projection}
To address these issues, we propose a type-shared feature projection to deal with the heterogeneity.
In HINs, nodes of different types typically reside in distinct feature spaces, which hinders the information aggregation~\cite{wang2019heterogeneous,yang2023simple}.
We assume that there exists a unified feature space $\mathcal{U}\in \mathbb{R}^{d_u}$. For each node type $t \in \mathcal{T}$, where $\mathcal{T}=\{\psi(v_i)|v_i\in V\}$, we project the original feature of each node $x[v_i]$ to the unified feature space $\mathcal{U}$ through a learnable type-shared linear layer:
\begin{equation}\label{eq:feature_transformation}
\setlength{\abovedisplayskip}{3pt}
\setlength{\belowdisplayskip}{3pt}
    \hat{x}[v_i] = W_t x[v_i]
\end{equation}
where $\forall v_i \in V$, $t$ represents the type of node $v_i$, i.e., $\psi(v_i)=t$, $W_t\in \mathbb{R}^{d_t \times d_u}$ is a learnable and type-specific linear layer, $d_u$ and $d_t$  denote the feature dimension of the unified space $\mathcal{U}$ and the original space of node type $t$, respectively.
Equation~\ref{eq:feature_transformation} enables nodes of different types to effectively aggregate information in a unified space.
It significantly alleviates the burden on systems or end-users as they no longer need to design predefined meta-paths or a series of relationship constraints.
It is worth noting that we still preserve the heterogeneity of edges in $\mathcal{H}_h$ rather than disregarding it, and propose an edge semantic attention mechanism (elaborated in Section~\ref{sec:information_integration}) to capture more fine-grained heterogeneous information.

\subsubsection{Query-dependent Graph}
In addition to the query-independent information inherent in HINs, query information is also very crucial for community search.
Therefore, we encode query-dependent information $\mathcal{X}_q$ into the HIN $\mathcal{H}$, denoted as $\mathcal{H}_q$, where the graph structure maintains consistency with $\mathcal{H}$.
Here, $\mathcal{X}_q \in \mathbb{R}^{n\times 1}$ is associated with the query node $q$ and is composed of $x_q[v_i]$, $\forall{v_i} \in V$.
\begin{equation}\label{eq:xq}
\setlength{\abovedisplayskip}{3pt}
\setlength{\belowdisplayskip}{3pt}
    x_q[v_i] = \begin{dcases}
    1 &\text{if } v_i = q; \\
    0 &\text{if } v_i \neq q.
    \end{dcases}
\end{equation}
The query information $\mathcal{X}_q$ indicates whether they are query nodes.
The same as $\mathcal{H}_h$, we also preserve the type of each edge in $\mathcal{H}_q$. This is done to preserve heterogeneous structural information around the query node during the information integration, rather than simply treating $\mathcal{H}_q$ as a homogeneous graph.

As shown in Figure~\ref{fig:framework}, we obtain $\mathcal{H}_h$ and $\mathcal{H}_q$ from $\mathcal{H}$ by employing type-shared feature projection and introducing query-dependent information.
In general, $\mathcal{H}_h$ and $\mathcal{H}_q$ offer two distinct views containing complementary information. The former describes the features of nodes with various types in a unified space, which is beneficial to information extraction and integration in the subsequent steps. The latter delineates the local structural information around the query node, which is crucial for community search.
\vspace{-3pt}

\subsection{Information Integration}
\label{sec:information_integration}
To better extract and integrate information from different views ($\mathcal{H}_h$ and $\mathcal{H}_q$), we designed two encoders: the heterogeneous encoder and the query encoder.
First, we propose the edge semantic attention mechanism to aggregate neighbor information, which is the basis of these two encoders.
\vspace{-3pt}

\subsubsection{Edge Semantic Attention Mechanism}
Different from homogeneous graphs, edges in HINs contain complex relationships, which are crucial for understanding the relationships between nodes and mining valuable community information.
Therefore, the heterogeneity inherent in edges still requires careful consideration.
Existing methods~\cite{fang2020effective,jian2020effective,jiang2022effective,zhou2023influential,liu2023significant} treat different types of relations equally when identifying communities. However, in multi-type communities, the contributions of different relations usually vary across different communities.
To address the above issues, we propose the edge semantic attention mechanism as follows:
\begin{equation}
\setlength{\abovedisplayskip}{3pt}
\setlength{\belowdisplayskip}{3pt}
    \alpha_{ij} = \frac{\exp{\Big(\mathrm{LeakyReLU}\big(\vec{\mathbf{a}}^T[\mathbf{W}\vec{\mathbf{h}}_i\|\mathbf{W}\vec{\mathbf{h}}_j\|\mathbf{W_e}\vec{\mathbf{e}}_{ij}]\big)\Big)}}{\sum_{r\in \mathcal{N}_i}\exp{\Big(\mathrm{LeakyReLU}\big(\vec{\mathbf{a}}^T[\mathbf{W}\vec{\mathbf{h}}_i\|\mathbf{W}\vec{\mathbf{h}}_r\|\mathbf{W_e}\vec{\mathbf{e}}_{ir}]\big)\Big)}},
    \label{eq:attention}
\end{equation}
where $\alpha_{ij}$ and $\vec{\mathbf{e}}_{ij}$ represent the attention weight and the edge features between $v_i$ and $v_j$ respectively, $\mathcal{N}_i$ denotes the neighbors of node $v_i$ in the graph $\mathcal{H}$, $\vec{\mathbf{h}}_i$ and $\vec{\mathbf{h}}_j$ are the representation vectors of node $v_i$ and $v_j$, and $\vec{\mathbf{a}}^T$, $\mathbf{W}$ and $\mathbf{W_e}$ are the learnable parameters.
Compared to the attention mechanisms solely based on nodes~\cite{velivckovic2017graph}, the edge semantic attention mechanism fully considers the heterogeneity of edges by encoding various relationships into attention weights, enriching node representations.
Furthermore, this mechanism empowers our model to dynamically assign weights to different edge types based on the specific communities, thereby preserving valuable information in diverse relationships.

In real-world HINs, not all networks have edge features. Therefore, we consider the following two scenarios:
the HINs include or lack edge features.
For the former, we can seamlessly apply equation~\ref{eq:attention}.
For the latter, to encode the information of edge types, we introduce an embedding matrix $\mathbf{E}$, which assigns a learnable feature vector to each edge type.
We formalize the embedding matrix: $\vec{\mathbf{e}}_{i,j} = \mathbf{E}[t, :]$, 
where $t \in \{1,2,\ldots,|\mathcal{E}|\}$ is a subscript of $\mathcal{E}= \{\phi(e_{i,j})|e_{i,j}\in E\}$, indicating the corresponding type of edge $e_{i,j}$, $\mathbf{E}\in \mathbb{R}^{|\mathcal{E}|\times d_e}$, $d_e$ denotes the feature dimension of edges.
\vspace{-4pt}



\subsubsection{Heterogeneous Encoder}
The heterogeneous encoder is dedicated to capturing query-independent information in $\mathcal{H}_h$.
Compared to previous studies in HINs~\cite{jian2020effective,fang2020effective,jiang2022effective,zhou2023influential}, the heterogeneous encoder dynamically considers the contributions of different relations to communities, rather than treating them equally.

The heterogeneous encoder is equipped with the edge semantic attention mechanism, and iteratively combines the neighbor information as follows:
\begin{equation}\label{eq:h_agg}
\setlength{\abovedisplayskip}{3pt}
\setlength{\belowdisplayskip}{3pt}
    \vec{\mathbf{h}}_{i}^{l} = \sigma_g\Big(\underset{k=1}{\overset{K}{||}} \sum_{j\in \mathcal{N}_i}\alpha_{ijk}^l \mathbf{W}_k^l\vec{\mathbf{h}}_j^{l-1} + \mathbf{W_r}^{l}\vec{\mathbf{h}}_i^{l-1}\Big),
\end{equation}
where $\vec{\mathbf{h}}_{i}^{l}$ denotes the representation of node $v_i$ in the $l$-th layer, $\alpha_{ijk}^l$ represents the $k$-th edge semantic attention calculated by Equation~\ref{eq:attention} for selective information aggregation, $||$ denotes the concatenation, and $\sigma_g$ is the $\text{ELU}$~\cite{velivckovic2017graph} activation function.
$\mathbf{W}_r^{l}\vec{\mathbf{h}}_i^{l-1}$ refers to the residual connection designed to prevent over-smoothing and gradient vanishing problems~\cite{li2018deeper,xu2018representation}.
When $l = 0$, $\vec{\mathbf{h}}_{i}^{0} = \hat{x}[v_i]$.

Through Equation~\ref{eq:h_agg}, we can uniformly consider structural information and node features with any dimension, avoiding ignoring node features or only considering low-dimensional features as in previous studies~\cite{jian2020effective,fang2021cohesive,zhou2023influential}.
Compared to ML-based community search methods~\cite{gao2021ics,DBLP:journals/pvldb/JiangRCHZH22,li2023coclep} in homogeneous graphs, our proposed heterogeneous encoder integrates edge types and dynamically considers the contributions of different relations instead of  treating them equally, which significantly enhances the understanding of complex community patterns in HINs.

\subsubsection{Query Encoder}
The query encoder is dedicated to obtaining query-dependent information in $\mathcal{H}_q$.
Compared to previous studies~\cite{jian2020effective,DBLP:journals/pvldb/JiangRCHZH22,zhou2023influential}, our proposed query encoder aggregates query information based on different types of edges to promote the model's understanding of complex relationships in HINs.
Specifically, the query encoder consists of multiple GNN layers, each incorporating the edge semantic attention mechanism as follows:
\begin{equation}\label{eq:hq_agg}
\setlength{\abovedisplayskip}{3pt}
\setlength{\belowdisplayskip}{3pt}
    \vec{\mathbf{h}}_{qi}^{l} = \sigma_g\Big(\underset{k=1}{\overset{K}{||}} \sum_{j\in \mathcal{N}_i}\alpha_{ijk}^l \mathbf{W}_k^l\vec{\mathbf{h}}_{qj}^{l-1} + \mathbf{W_r}^{l}\vec{\mathbf{h}}_{qi}^{l-1}\Big),
\end{equation}
where $\vec{\mathbf{h}}_{qi}^{l}$ denotes the query-dependent representation of node $v_i$ in the $l$-th layer, and $\vec{\mathbf{h}}_{qi}^{0}=x_q[v_i]$.
Note that for brevity, the learnable parameters, such as $\alpha_{ijk}^l$, $\mathbf{W}_k^l$ and $\mathbf{W_r}^{l}$, are no longer distinguished in Equations~\ref{eq:h_agg} and~\ref{eq:hq_agg}, but they train these parameters separately.
The difference between Equation~\ref{eq:h_agg} and Equation~\ref{eq:hq_agg} lies in their focus on different views, dealing with query-independent and query-dependent information, respectively.
Through query encoder, we can effectively capture the local structural information around query nodes to facilitate community search.

\subsubsection{Feature Fusion}
The above heterogeneous encoder and query encoder respectively extract information from different views ($\mathcal{H}_h$ and $\mathcal{H}_q$).
As the importance of each view may vary for each node, we cannot simply treat these different information equally.
Therefore, we introduce a feature fusion module based on the attention mechanism after each layer to take full advantage of complementary information from different views.
Specifically, node $v_i$ at layer $l$ has two different representations ($\vec{\mathbf{h}}_{i}^{l}$ and $\vec{\mathbf{h}}_{qi}^{l}$).
Then, we use an attention mechanism to fuse these complementary representations for node $v_i$, as follows:
\begin{equation}\label{eq:fusion}
\setlength{\abovedisplayskip}{3pt}
\setlength{\belowdisplayskip}{3pt}
\begin{aligned}
    e_{i}^l=\vec{\mathbf{u}}^{\intercal} \cdot \vec{\mathbf{h}}_{i}^{l}, &\quad e_{qi}^l=\vec{\mathbf{u}}_q^{\intercal} \cdot \vec{\mathbf{h}}_{qi}^{l}, \\
    \beta_{i}^l=\frac{\exp \left(e_{i}^l\right)}{\exp \left(e_{i}^l\right)  + \exp \left(e_{qi}^l\right)}, &\quad \beta_{qi}^l=\frac{\exp \left(e_{qi}^l\right)}{\exp \left(e_{i}^l\right)  + \exp \left(e_{qi}^l\right)}, \\
    \vec{\mathbf{h}}_{fi}^{l} = \beta_{i}^l \cdot &\vec{\mathbf{h}}_{i}^{l} + \beta_{qi}^l \cdot \vec{\mathbf{h}}_{qi}^{l},
\end{aligned}
\end{equation}
where $\vec{\mathbf{h}}_{fi}^{l}$ is the fused node representation after $l$ layer, and $\vec{\mathbf{u}}^{\intercal}$ and $\vec{\mathbf{u}}_q^{\intercal}$ are the parameterized attention vectors, respectively. $\beta_{i}^l$ and $\beta_{qi}^l$ can be interpreted as the importance of different views for the node.
As shown in Figure~\ref{fig:framework}, for the query encoder, we use the fused node representation as the input for the next layer.
For the heterogeneous encoder, we only use the output of the heterogeneous encoder itself in the $l$-th layer as the input of the $(l+1)$-th layer.
In this way, we make the heterogeneous encoder independent of the query information, which provides stable prior knowledge about the HINs.

Furthermore, we use the output of the feature fusion module in the final layer as the node representation, which is utilized to compute the probability of each node belonging to a community:
\begin{equation}\label{eq:probability}
\setlength{\abovedisplayskip}{3pt}
\setlength{\belowdisplayskip}{3pt}
    p_{i} = \sigma\Big(\text{MLP}(\vec{\mathbf{h}}_{fi}^{L})\Big).
\end{equation}
where $\sigma$ indicates the Sigmoid function.

\subsection{Depth-based Heuristic Search Algorithm}
When dealing with large-scale graphs, traditional community search methods~\cite{jiang2022effective,liu2023significant} usually improve query efficiency by designing appropriate indexes.
However, ML-based community search methods~\cite{gao2021ics,DBLP:journals/pvldb/JiangRCHZH22} face significant challenges due to the substantial computational overhead.

To address this issue, we propose the depth-based heuristic search strategy to improve the query efficiency in the large-scale graphs.
The core hypothesis is that: the nodes closer to the query node are more likely to belong to the same community.
This assumption is grounded in real-life observations: individuals within a community tend to have stronger connections with each other compared to individuals outside the community.
In other words, nodes within the same community typically have more edges and shorter paths between them.
Therefore, we propose the depth-based heuristic strategy to limit the exploration depth, reducing the search on distant nodes and thereby enhancing search efficiency.

\begin{algorithm}[t]
\caption{Depth-based heuristic search algorithm}
\label{alg:multi_search_p}
\LinesNumbered 
\KwIn{User-specified node types $\mathcal{S}_{\mathcal{A}}$ in the target community, query node $q$, node probability $p_i$, $\mathcal{H}=(V, E, \psi, \phi)$, threshold $\gamma$, maximum depth $d_{max}$}
\KwOut{The target community $\mathcal{C}_q$}
Initialize $\mathcal{Q}\leftarrow$ empty queue, $\mathcal{C}_q \leftarrow \{q\}$\;
$\mathcal{V} \leftarrow \varnothing$ \tcp{recorded visited nodes}
$\mathcal{Q} \leftarrow (q, 0)$ \tcp{Add $q$ and depth 0 into the queue}

\While{$\mathcal{Q}$ is not empty}{
    (node $v_i$, depth $d$) $\gets$ dequeue from $\mathcal{Q}$\;
    \If{$v_i \notin \mathcal{V}$ and $d<d_{max}$}{
        $\mathcal{V} \gets \mathcal{V} \cup \{v_i\}$ \tcp{add visited nodes}
        \ForEach{$v_j \in \mathcal{N}_i$}{
            $\mathcal{Q} \leftarrow (v_j, d+1)$ \tcp{add new node for BFS}
            \If{$\psi(v_j) \in \mathcal{S}_{\mathcal{A}}$ and $p_j > \gamma$}{\
                $\mathcal{C}_q \gets \mathcal{C}_q \cup \{v_j\}$\;
            }
        }
    }
}
\Return The target community $\mathcal{C}_q$.
\end{algorithm}

Algorithm~\ref{alg:multi_search_p} outlines the depth-based heuristic search algorithm.
The search algorithm is designed to identify the target community based on user-specified node types $\mathcal{S}_{\mathcal{A}}$ and the probability of each node belonging to the community calculated by Equation~\ref{eq:probability}.
It is worth noting that, for the purpose of subsequent efficiency and performance comparisons, we still employ BFS as the search algorithm for FCS-HGNN.
First, we initialize a set $\mathcal{V}$ to record visited nodes, a queue $\mathcal{Q}$ to store node information and record the depth of each node from the query node $q$ (Lines 1-3).
Then, we use the depth to determine whether to explore the node (Line 6).
In the loop, for each neighbor $v_j$ of $v_i$, we do not restrict the criteria for adding $v_j$ to the queue $\mathcal{Q}$, because the neighbors of low probability ($p_j < \gamma$) or non-target type nodes ($\psi(v_j) \notin \mathcal{S}_{\mathcal{A}}$) may also be community members (Line 9).
For the node that satisfies the community conditions, we add it to the target community $\mathcal{C}_q$ (Lines 10-11).
Through Algorithm~\ref{alg:multi_search_p}, we identify target communities based on the node probability while ensuring the connectivity of the communities.

\subsection{Training and Online Query}

\subsubsection{Neighbor Sampling}

To improve the training efficiency especially for large-scale graphs, we introduce a neighbor sampling algorithm, which recursively samples a fixed number of neighbors for each node to form a subgraph for training~\cite{hamilton2017inductive}.
Algorithm~\ref{alg:neighbor_sampler} outlines the neighbor sampling algorithm, where $S_l$ represents the set of nodes sampled in the $l$-th layer, $N_{v_{i}}^{l-1}$ represents the set of nodes sampled from the neighbors of $v_i$, utilized for training in the $(l-1)$-th layer, and fanouts $\{f_1, f_2, ..., f_L\}$ are hyperparameters indicating the number of neighbors that should be sampled at each layer.
We initialize $S_L$ with the labeled nodes (Line 1), as the probabilities of these nodes are ultimately used to compute the loss in Equation~\ref{eq:loss}.
Then, for each node $v_i$ in $S_l$, we sample a fixed number $f_{l-1}$ of nodes from the neighbors of $v_i$, and add them to $S_{l-1}$ (Lines 2-6).
Finally, we extract the corresponding subgraph $\mathcal{H}'$ from $\mathcal{H}$ based on the nodes sampled at each layer (Line 7).

\begin{algorithm}[htbp]
\caption{Neighbor Sampling of LS-FCS-HGNN}
\label{alg:neighbor_sampler}
\LinesNumbered 
\KwIn{labeled Nodes ($q$, $pos(q)$ and $neg(q)$), $\mathcal{H}$, the number of layers $L$, Fanouts $f_1, f_2, ..., f_L$}
\KwOut{The subgraph for training}
 $S_{L} \leftarrow \{q, pos(q), neg(q)\}$\tcp{Initialization}
 \For{$l=L$ to $1$}{
  $S_{l-1} \leftarrow \varnothing$ \tcp{$S_{l-1}$ record the nodes sampled in the $(l-1)$-th layer}
  \ForEach{node $v_i$ in $S_{l}$}{
   $N_{v_{i}}^{l-1} \leftarrow$ Sample $f_{l-1}$ neighbors of $v_i$ from $\mathcal{H}$\;
   $S_{l-1} \leftarrow S_{l-1} \cup \{N_{v_{i}}^{l-1}\}$\;
  }
 }
 Extract the subgraph $\mathcal{H}'$ of $\mathcal{H}$ with $\bigcup_{l=0}^{L} S_l$\;
 \Return The subgraph $\mathcal{H}'$
\end{algorithm}

\subsubsection{Training}
For the training stage, following previous work~\cite{gao2021ics,DBLP:journals/pvldb/JiangRCHZH22}, we define the community search problem as a binary node classification problem, where the goal is to train the model to determine whether each node belongs to the community under query $q$.
We denote $y \in \{0,1\}$ as the true label of each node under query $q$.
The label of $y_i$ will change with different queries $q$.
Then, we utilize the Binary Cross Entropy (BCE) function as the loss function:
\begin{equation}\label{eq:loss}
\setlength{\abovedisplayskip}{3pt}
\setlength{\belowdisplayskip}{3pt}
    \mathcal{L} = -\sum_{q\in \mathcal{D}} \frac{1}{n}\sum_{i=1}^n\Big(y_{i}\log (p_{i}) + (1-y_{i})\log (1-p_{i})\Big),
\end{equation}

\begin{algorithm}[htbp]
\caption{The training process of (LS-)FCS-HGNN}
\label{alg:training}
\LinesNumbered 
\KwIn{Training data $\mathcal{D}$, HIN $\mathcal{H}$, the number of multi-head attention $K$, the number of layer $L$}
\KwOut{The optimal parameters of FCS-HGNN}
Initialize the parameters of FCS-HGNN\; 
\While{not converge}{
  \For{$(q, pos(q), neg(q)) \in \mathcal{D}$}{
    \tcc{\textcolor{black}{Special For LS-FCS-HGNN}}
      \tcp{Neighbor Sampling}
        $\mathcal{H'} \leftarrow$ Algorithm~\ref{alg:neighbor_sampler} ($\{q, pos(q), neg(q)\}$, $\mathcal{H}$, $L$, $f_1, f_2, ..., f_L$)\;
    \tcc{\textcolor{black}{Common Part}}
    \tcp{Heterogeneous \& Query Information Preprocessing}
    $\mathcal{H}_h \leftarrow$ type-shared feature projection by Eq.~\ref{eq:feature_transformation}\;
    $\mathcal{H}_q \leftarrow$ query-dependent information by Eq.~\ref{eq:xq}\;
    \tcp{Information Integration}
    $\vec{\mathbf{h}}^{1} \leftarrow \text{\textit{Hetero. Encoder}}^1(\mathcal{H}_h,\vec{\mathbf{h}}^{0})$ in Eq.~\ref{eq:h_agg}\;
    $\vec{\mathbf{h}}_q^{1} \leftarrow \text{\textit{Query Encoder}}^1(\mathcal{H}_q,\vec{\mathbf{h}}_{q}^{0})$ in Eq.~\ref{eq:hq_agg}\;
    $\vec{\mathbf{h}}_{f}^{1} \leftarrow \text{\textit{Feature Fusion}}^1(\vec{\mathbf{h}}^{1},\vec{\mathbf{h}}_{q}^{1})$ in Eq.~\ref{eq:fusion}\;
    $l = 1$\;
    \While{$l<L$}{
        $\vec{\mathbf{h}}^{l+1} \leftarrow \text{\textit{Hetero. Encoder}}^{l+1}(\mathcal{H},\vec{\mathbf{h}}^{l})$ in Eq.~\ref{eq:h_agg}\;
        $\vec{\mathbf{h}}_{q}^{l+1} \leftarrow \text{\textit{Query Encoder}}^{l+1}(\mathcal{H}_q,\vec{\mathbf{h}}_{f}^{l})$ in Eq.~\ref{eq:hq_agg}\;
        $\vec{\mathbf{h}}_{f}^{l+1} \leftarrow \text{\textit{Feature Fusion}}^{l+1}(\vec{\mathbf{h}}^{l+1},\vec{\mathbf{h}}_{q}^{l+1})$ in Eq.~\ref{eq:fusion}\;
        $l = l + 1$\;
    }
    $p \leftarrow$ calculated probability of each node by Eq.~\ref{eq:probability}\;
    Update the parameters according to the loss in Eq.~\ref{eq:loss}\;
  }
}
\Return The optimal parameters of (LS-)FCS-HGNN.
\end{algorithm}

Algorithm~\ref{alg:training} outlines the training process of both our proposed FCS-HGNN and LS-FCS-HGNN.
It is noteworthy that $\mathcal{H}_h$ and $\mathcal{H}_q$ differ solely in terms of node features, sharing the same graph structure, which ensuring that there is no additional computational overhead.
Compared to FCS-HGNN, LS-FCS-HGNN first samples the subgraph $\mathcal{H}'$ for training (Line 4).
Furthermore, we perform neighbor sampling at each epoch to obtain a variety of subgraphs, which allows the model to observe various structures and thereby comprehensively understand the entire community pattern.
In this way, the neighbor sampling strategy effectively reduces the computational overhead of community search in large-scale graphs while maintaining the community effectiveness.
Subsequent experiments in Section~\ref{sec:cs_results} further demonstrate the effectiveness and efficiency of the neighbor sampling strategy.


\subsubsection{Online Query}
For the online query phase, we utilize the best model obtained during the training phase to execute queries, instead of retraining the model for each query~\cite{gao2021ics}.
For a new query $q_{new}$, we first employ the model to calculate the probability of each node belonging to the community.
Then, we output the target community based on the search algorithm, as shown in Algorithm~\ref{alg:multi_search_p}.

\subsection{Time Complexity}

\begin{theorem}\label{the:training_FCS-HGNN}
Let $T$ denote the number of training iterations, $\vert V\vert$ represent the number of nodes, $\vert E\vert$ represent the number of edges, $\vert \mathcal{D}\vert$ be the quantity of the training set, and $F$ and $F'$ represent the feature dimensions before and after linear transformation, respectively.
The time complexity for a layer of heterogeneous encoder or query encoder is $\mathcal{O}(\vert V\vert F F' + \vert E\vert F')$.
For the FCS-HGNN with $L$ layers, the time complexity for each iteration would be $\mathcal{O}(L(\vert V\vert F F' + \vert E\vert F'))$.
Thus, \textbf{the time complexity of the training process of FCS-HGNN is $\mathcal{O}(TL\vert \mathcal{D}\vert (\vert V\vert F F' + \vert E\vert F'))$}, which primarily depends on the size of the graph, such as $\vert V\vert$ and $\vert E\vert$, and
\textbf{the time complexity of the training process of LS-FCS-HGNN is $\mathcal{O}(TL\vert \mathcal{D}\vert (n F F' + e F' + Ln))$}, where $Ln$ is the time complexity of neighbor sampling, $n=\sum_{l=1}^L n_l$.
\end{theorem}

\begin{theorem}
    The time complexity of depth-based heuristic search algorithm (Algorithm~\ref{alg:multi_search_p}) is bounded by $\mathcal{O}(\vert V\vert + \vert E\vert)$, which depends on the maximum exploration depth $d_{max}$ as well as the graph structure.
    Thus, \textbf{the time complexity of the query process of LS-FCS-HGNN is bounded by} $\mathcal{O}(L(n F F' + e F' + Ln) + \vert V\vert + \vert E\vert)$, and
    \textbf{the time complexity of the query process for FCS-HGNN is $\mathcal{O}(L(\vert V\vert F F' + \vert E\vert F') + \vert V\vert + \vert E\vert)$}.
\end{theorem}

Since $n$ and $e$ are much smaller than $\vert V \vert$ and $\vert E \vert$, the training and query efficiency of LS-FCS-HGNN will be significantly higher than that of FCS-HGNN.
In the next section, we compare the time performance of our proposed algorithm with state-of-the-art methods.

\section{EXPERIMENTS}\label{section:experiment}
\subsection{Experimental Setup}

\subsubsection{Datasets.}
To comprehensively evaluate the performance of our proposed method, we conducted experiments on five real-world HINs, i.e., IMDB~\cite{luo2021detecting}, DBLP~\cite{fu2020magnn}, ACM~\cite{luo2021detecting}, Freebase~\cite{lv2021we}, and OGB-MAG~\cite{hu2020open}.
Table~\ref{tab:datasets} reports the statistics of the datasets, where $\vert \mathcal{A} \vert$ and $\vert \mathcal{E} \vert$ denote the number of node types and edge types, respectively.
Node features are necessary for all datasets.
Notably, Freebase possesses a wealth of edge relationships, while OGB-MAG is a large-scale HIN derived from the OGB benchmark~\cite{hu2020open}.


\begin{table}[]
\centering
\caption{Dataset statistics}
    \resizebox{.7\linewidth}{!}{
    \begin{tabular}{@{}ccccc@{}}
    \toprule
    \textbf{Dataset} & \textbf{$\vert V \vert$} & \textbf{$\vert E \vert$} & \textbf{$\vert \mathcal{A} \vert$} & \textbf{$\vert \mathcal{E} \vert$} \\ \midrule
    IMDB             & 19,103                   & 74,880                   & 4                                  & 6                                  \\
    DBLP             & 26,128                   & 265,694                  & 4                                  & 6                                  \\
    ACM              & 31,807                   & 160,321                  & 4                                  & 6                                  \\
    Freebase         & 180,098                  & 1,057,688                & 8                                  & 36                                 \\
    OGB-MAG          & 1,939,743                & 36,805,743               & 4                                  & 7                                  \\ \bottomrule
    \end{tabular}
}
\vspace{-8pt}
\label{tab:datasets}
\end{table}

\subsubsection{Community Generation.}
In reality, it is challenging to manually collect labeled communities for training, especially multi-type communities in large-scale graphs.
Thus, we adopt traditional community search methods in HINs to automatically label the single-type~\cite{jiang2022effective} or multi-type communities~\cite{jian2020effective} as the ground truth. 
Specifically, the objective of~\cite{jiang2022effective} is to identify $(k,\Psi)$-NMC communities containing nodes of a single type, while~\cite{jian2020effective} aims to search for relational communities containing nodes of different types.
Since other traditional community search methods~\cite{fang2020effective,zhou2023influential} in HINs focus on different community patterns that are not consistent with those of~\cite{jian2020effective} and~\cite{jiang2022effective}, it would be unfair to compare these methods.
ML-based methods do not rely on predefined community patterns. Instead, they learn community patterns in a data-driven manner.
Therefore, although these communities~\cite{jian2020effective,jiang2022effective} impose strict constraints on community patterns, they provide a relatively fair comparison for ML-based methods in the absence of labeled communities in HINs.
For these reasons, we mainly compare with state-of-the-art ML-based methods.
For the single-type community, we designate one type of node as the primary node.
For the DBLP dataset, we consider either author or paper nodes as the primary nodes, resulting in two variants: DBLP-A and DBLP-P, respectively.
For the other datasets, we designate the primary node for single-type community according to~\cite{luo2021detecting}.
For the multi-type community, we set the paper and author nodes as primary nodes for DBLP, ACM, and OGB-MAG, movie and actor nodes as primary nodes for IMDB, film and people nodes as primary nodes for Freebase.

\subsubsection{Baselines.}
To our knowledge, we are the first to propose the multi-type community search method based on graph neural networks in HINs, thus there are no direct studies available for comparison.
Therefore, we compare our proposed FCS-HGNN and LS-FCS-HGNN with state-of-the-art methods in homogeneous graphs, including ICS-GNN~\cite{gao2021ics}, QD-GNN~\cite{DBLP:journals/pvldb/JiangRCHZH22}, COCLE and COCLEP~\cite{li2023coclep}.
Specifically, we transform HINs into homogeneous graphs using meta-paths~\cite{wang2019heterogeneous,zhang2019heterogeneous,luo2021detecting} and augment these baselines with our proposed type-shared feature projection to enable these methods to identify multi-type communities.
Other methods~\cite{hashemi2023cs,behrouz2022cs,fang2023community} may face challenges in being applied to HINs even with preprocessing, therefore we do not consider comparing with them in this paper.

\subsubsection{Evaluation Metrics.}
To evaluate the quality of the identified communities, we employed three metrics: F1-score~\cite{fang2023community}, Jaccard Similarity~\cite{zhang2020seal}, and Normalized Mutual Information (NMI)~\cite{danon2005comparing}.

\subsubsection{Implementation Details.}
For our model, we construct $L=2$ layers with dimensions of 64 or 128, and used Adam~\cite{kingma2014adam} for optimization. As for other hyperparameters, we set the learning rate to 0.001, the regularization parameter to 1e-4, the dropout rate to 0.5, the number of attention heads $K$ to 8, and the fanouts to $f_1 = 20, f_2 = 10$.
The threshold $\gamma$ is obtained by searching for the best F1 score from the validation set.
For LS-FCS-HGNN, we traverse the maximum depth $d_{max}$ from 2 to 10 and record the best results.
For all the baselines, we fine-tuned their parameters according to their original papers and recorded their best performance.

\subsection{Community Search Performance}
\label{sec:cs_results}

\begin{table*}[ht]
\centering
\caption{Performance comparison in \textbf{single-type} community search (in percentage).}
    \resizebox{\linewidth}{!}{
\begin{tabular}{@{}ccccccccccccccccccc@{}}
\toprule
\multirow{2}{*}{Method}      & \multicolumn{3}{c}{DBLP-A}                                & \multicolumn{3}{c}{DBLP-P}                                & \multicolumn{3}{c}{ACM}                                   & \multicolumn{3}{c}{IMDB}                                  & \multicolumn{3}{c}{Freebase}                              & \multicolumn{3}{c}{OGB-MAG}                              \\ 
            & F1                & JS                & NMI               & F1                & JS                & NMI               & F1                & JS                & NMI               & F1                & JS                & NMI               & F1                & JS                & NMI               & F1                & JS                & NMI              \\ \midrule
ICS-GNN     & 58.38             & 46.94             & \underline{37.44} & \underline{73.75} & \underline{68.88} & \underline{69.16} & 42.48             & 31.22             & 25.23             & 31.2              & 19.42             & 8.21              & 46.44             & \underline{43.82} & \underline{45.07} & o.o.m             & o.o.m             & o.o.m            \\
QD-GNN      & 54.43             & 42.84             & 27.34             & 66.81             & 65.42             & 64.75             & 40.43             & 26.33             & 21.39             & \underline{61.01} & \underline{44.46} & 10.47             & \underline{47.48} & 39.2              & 37.58             & o.o.m             & o.o.m             & o.o.m            \\
COCLE       & 57.70              & 43.63             & 31.59             & 68.12             & 62.02             & 67.99             & 40.69             & 26.98             & 34.04             & 60.47             & 44.19             & \underline{13.44} & 38.95             & 33.14             & 36.00                & o.o.m             & o.o.m             & o.o.m            \\
COCLEP      & \underline{63.41} & \underline{48.64} & 35.05             & 66.69             & 59.23             & 60.89             & \underline{55.76} & \underline{41.74} & \underline{38.35} & 50.93             & 34.53             & 10.53             & 43.8              & 38.19             & 36.62             & \underline{17.85} & \underline{11.33} & \underline{9.71} \\
FCS-HGNN    & \textbf{77.45}    & \textbf{64.38}    & \textbf{62.28}    & 78.99             & 74.23             & 74.59             & 76.73             & 75.38             & 75.52             & 67.35             & 51.07             & 42.29             & 54.44             & 51.82             & 53.48             & o.o.m             & o.o.m             & o.o.m            \\
LS-FCS-HGNN & 75.69             & 63.65             & 62.27             & \textbf{81.30}    & \textbf{76.35}    & \textbf{76.56}    & \textbf{79.39}    & \textbf{76.32}    & \textbf{75.81}    & \textbf{68.70}    & \textbf{52.61}    & \textbf{42.48}    & \textbf{54.44}    & \textbf{51.82}    & \textbf{53.48}    & \textbf{24.70}    & \textbf{15.90}    & \textbf{11.75}   \\ \bottomrule
\end{tabular}
}
\label{tab:single_type}
\vspace{-6pt}
\end{table*}

\begin{table*}[ht]
\centering
\caption{Performance comparison in \textbf{multi-type} community search.}
    \resizebox{\linewidth}{!}{
\begin{tabular}{@{}cccccccccccccccc@{}}
\toprule
\multirow{2}{*}{Method}      & \multicolumn{3}{c}{DBLP}                                  & \multicolumn{3}{c}{ACM}                                   & \multicolumn{3}{c}{IMDB}                                  & \multicolumn{3}{c}{Freebase}                              & \multicolumn{3}{c}{OGB-MAG}                             \\ 
            & F1                & JS                & NMI               & F1                & JS                & NMI               & F1                & JS                & NMI               & F1                & JS                & NMI               & F1                & JS               & NMI              \\ \midrule
ICS-GNN     & 40.40             & 27.22             & 35.02             & 37.48             & 30.68             & 34.14             & 35.82             & 28.41             & 22.04             & 31.47             & 22.00             & 26.86             & o.o.m             & o.o.m            & o.o.m            \\
QD-GNN      & 46.78             & 39.57             & 42.74             & 38.23             & \underline{33.69} & \underline{35.28} & 39.55             & \underline{33.93} & 28.45             & 34.96             & 27.10             & 26.38             & o.o.m             & o.o.m            & o.o.m            \\
COCLE       & \underline{52.82} & \underline{41.67} & \underline{45.23} & \underline{41.99} & 28.89             & 33.19             & \underline{41.39} & 32.68             & \underline{35.31} & 33.16             & 21.70             & 22.15             & o.o.m             & o.o.m            & o.o.m            \\
COCLEP      & 51.30             & 40.87             & 45.14             & 37.13             & 27.49             & 31.67             & 36.85             & 29.27             & 32.88             & \underline{56.71} & \underline{46.27} & \underline{47.64} & \underline{11.73} & \underline{9.42} & \underline{5.77} \\
FCS-HGNN    & \textbf{55.45}    & \textbf{42.13}    & \textbf{47.48}    & \textbf{53.01}    & \textbf{40.88}    & \textbf{46.31}    & 44.11             & 36.80             & 40.64             & 68.07             & 54.50             & 57.78             & o.o.m             & o.o.m            & o.o.m            \\
LS-FCS-HGNN & 55.19             & 41.87             & 47.15             & 52.67             & 40.43             & 45.86             & \textbf{44.15}    & \textbf{36.82}    & \textbf{40.69}    & \textbf{89.00}    & \textbf{84.54}    & \textbf{85.39}    & \textbf{19.07}    & \textbf{11.85}   & \textbf{8.45}    \\ \bottomrule
\end{tabular}
}
\label{tab:multi_type}
\vspace{-6pt}
\end{table*}

\subsubsection{Single-type Community Search.}
Since the single-type community is a special case of the multi-type community, we first compare the effectiveness of our proposed algorithms in the context of the single-type community search.
Table~\ref{tab:single_type} displays the experimental results of single-type community search in HINs.

From Table~\ref{tab:single_type}, it is evident that our proposed methods outperform other state-of-the-art methods across all datasets, achieving average improvements of 11.08\%/13.08\%/18.68\% on the F1/JS/NMI metrics, respectively.
This demonstrates the effectiveness of our methods when identifying single-type community search in HINs.
We observe that these baselines generally perform poorly on the NMI metric, indicating that they fail to capture the structure and node distribution of the target communities adequately.
Our method not only significantly surpasses the baselines on the NMI metric but also does not exhibit lopsided performance across the three metrics.
This indicates that our proposed methods effectively learn the complex community structure in HINs, which is crucial for community search.
Furthermore, it is illuminating to observe that LS-FCS-HGNN produced superior results in most datasets except for DBLP-A, demonstrating the effectiveness of the neighbor sampling and depth-based heuristic strategies.
Under the guidance of the depth-based heuristic strategy, LS-FCS-HGNN eliminates the interference from distant nodes, achieving superior results compared to FCS-HGNN which employs a full-graph search.

\subsubsection{Multi-type Community Search.}
Then, we evaluate the effectiveness of our proposed methods in the context of multi-type community search.
Table~\ref{tab:multi_type} presents the experimental results of multi-type community search in HINs.
It is worth emphasizing that for our proposed methods, the transition from single-type community search to multi-type community search only requires modifying the user-specific node types $\mathcal{S}_{\mathcal{A}}$ of their desired communities, which is very easy for users.
Conversely, we undertake a considerable amount of work to adapt these baselines for multi-type community search.
There is no need for any changes to the graphs or models for our methods, which significantly improves the flexibility and applicability of the algorithm.

As shown in Table~\ref{tab:multi_type}, our proposed methods exhibit outstanding performance across all datasets, achieving average improvements of 11.21\%/10.24\%/11.81\% on the F1/JS/NMI metrics, respectively.
On one hand, the type-shared feature projection adeptly addresses the heterogeneity of nodes in the absence of meta-paths or relational constraints, which maps nodes from different feature spaces into a unified feature space, thereby enhancing information integration.
On the other hand, the edge semantic attention mechanism selectively aggregates information from neighbors based on different relations, effectively leveraging fine-grained heterogeneous information and facilitating the model's understanding of HINs.
Our subsequent ablation experiments further confirmed this observation.
Moreover, we are able to obtain similar conclusions to single-type community search: the neighbor sampling and depth-based heuristic search strategies not only empower LS-FCS-HGNN to efficiently identify communities in large-scale graphs, but also improve the quality and precision of community identification.
This further demonstrates the effectiveness of our methods.


\subsubsection{Training Efficiency.}\label{sec:training_efficiency}
Figure~\ref{fig:training_time} illustrates the training time per epoch for various methods on all datasets.
As ICS-GNN requires retraining for each query, we did not record its result of training efficiency.
On the OGB-MAG dataset, QD-GNN, COCLE, and FCS-HGNN run out of GPU memory.
As shown in Figure~\ref{fig:training_time}, we observe that when the graph size is small, the training efficiency of each method is similar. As the size increases, the differences in training efficiency become more apparent.
In general, FCS-HGNN exhibits better training efficiency than COCLE and COCLEP.
However, our proposed LS-FCS-HGNN demonstrates exceptional training efficiency, surpassing all competing methods. Especially as the graph size increases, the advantage of LS-FCS-HGNN's training efficiency becomes increasingly significant.

\subsubsection{Query Efficiency.}
\begin{figure}[t]
    \begin{minipage}[c]{0.8\linewidth}
        \centering

        \begin{subfigure}[b]{\linewidth}
            \centering
            \includegraphics[width=\linewidth]{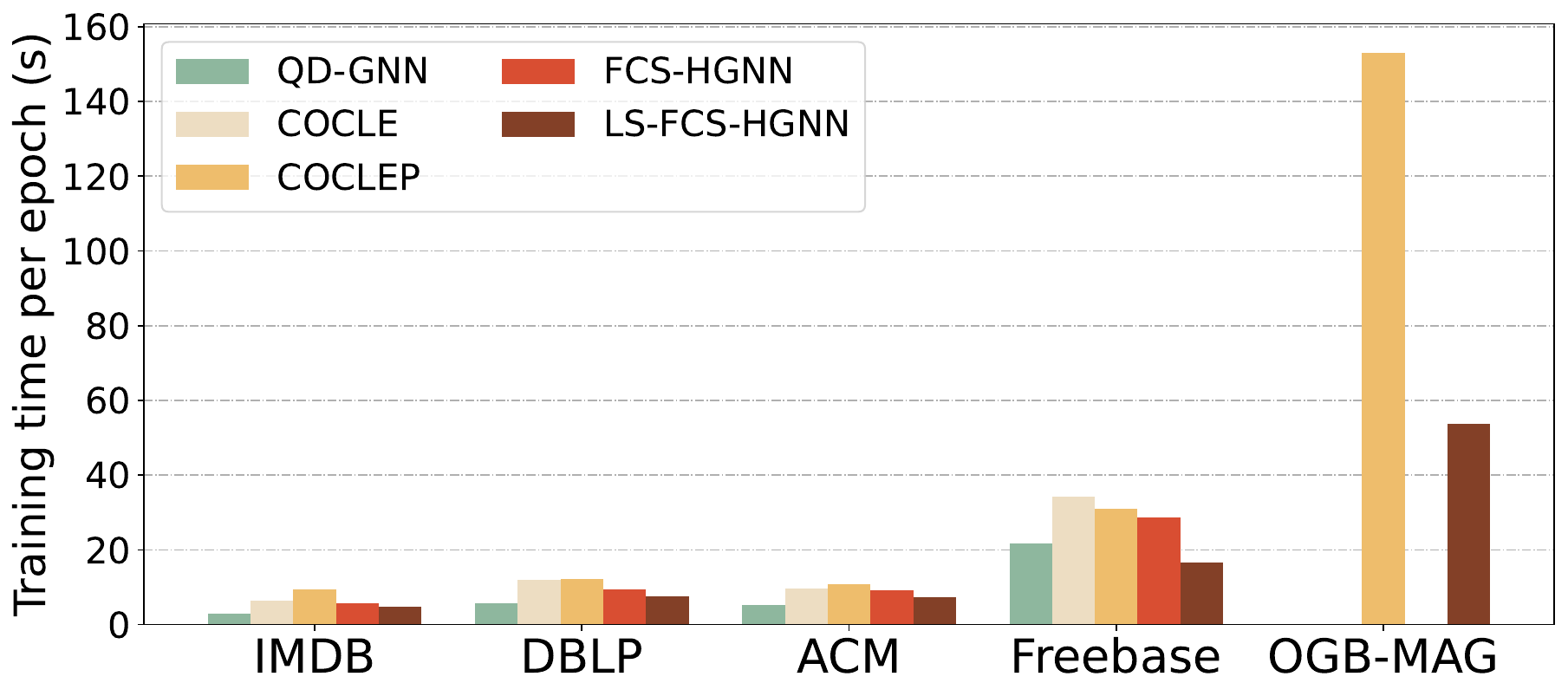} 
            \caption{Training Efficiency.}
    \label{fig:training_time}
        \end{subfigure}
        \begin{subfigure}[b]{\linewidth}
            \centering
            \includegraphics[width=\linewidth]{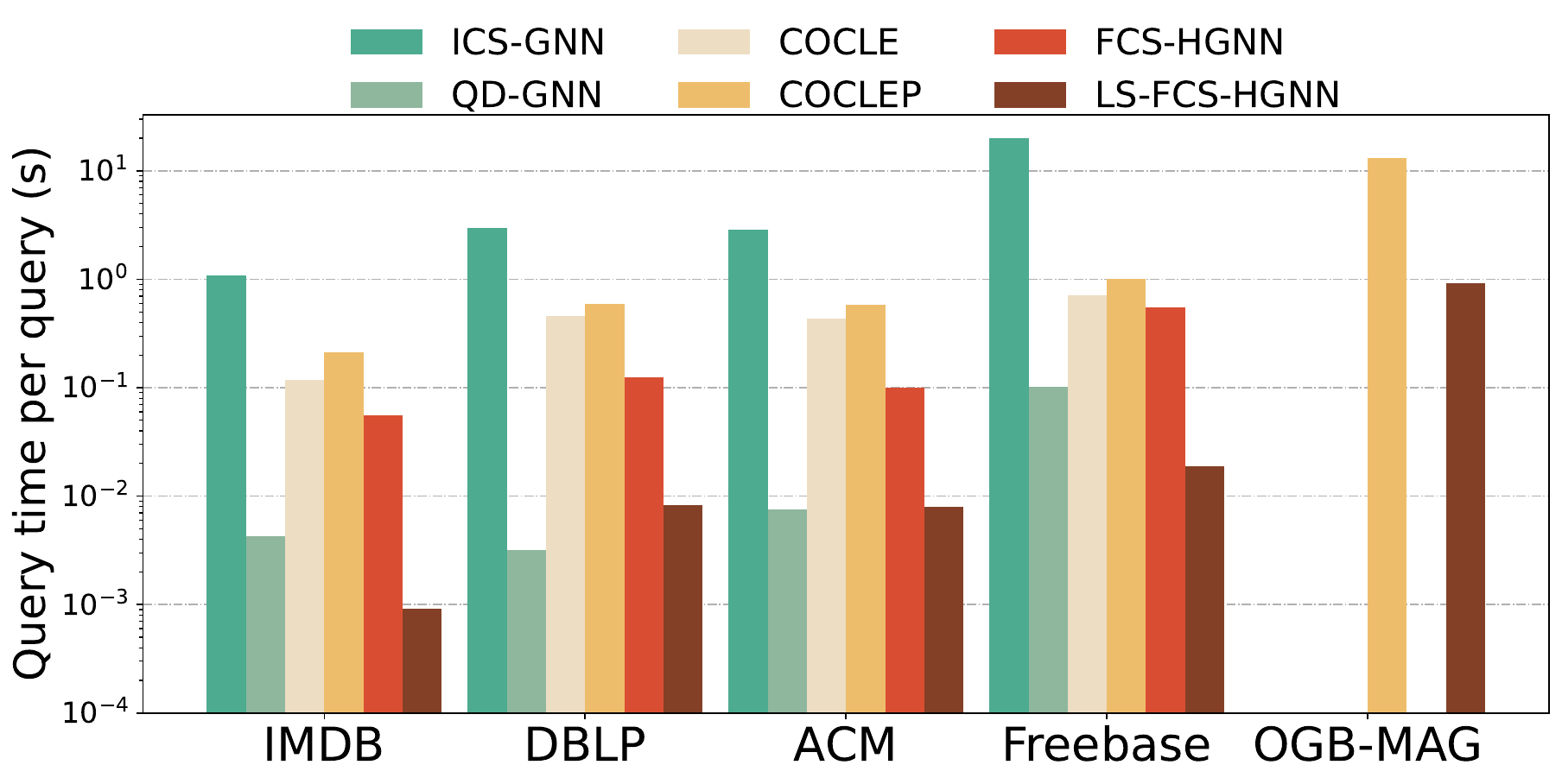} 
            \caption{Query Efficiency.} 
            \label{fig:inference_time}
        \end{subfigure}
        \caption{Efficiency comparison}
    \end{minipage}%
    \vspace{-2em}
\end{figure}

In addition to training efficiency, we also thoroughly compared the query efficiency of each method, as shown in Figure~\ref{fig:inference_time}.
It is worth noting that each grid on the y-axis of Figure~\ref{fig:inference_time} differs by a factor of 10.
Since ICS-GNN retrains the whole model for each new query, it exhibits the lowest query efficiency.
Similar to training efficiency, the query efficiency of FCS-HGNN significantly outperforms that of COCLE and COCLEP.
However, with the assistance of the depth-based heuristic search strategy, the query efficiency of LS-FCS-HGNN is significantly superior to all other methods.
As the graph size continues to increase, this advantage becomes even more pronounced.
This further demonstrates the effectiveness of our proposed depth-based heuristic search strategy. 




\subsection{Ablation Study}

\begin{figure*}
    \begin{minipage}[c]{0.33\linewidth}
        \centering

        \begin{subfigure}[b]{0.49\linewidth}
            \centering
            \includegraphics[width=\linewidth]{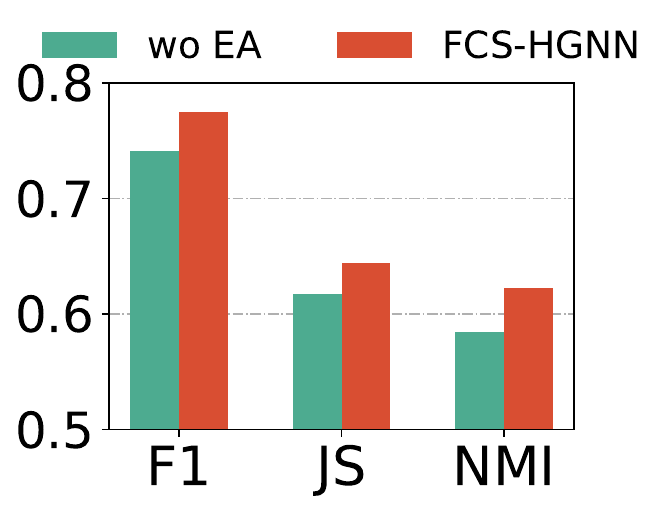} 
            \caption{DBLP}
        \end{subfigure}
        \begin{subfigure}[b]{0.49\linewidth}
            \centering
            \includegraphics[width=\linewidth]{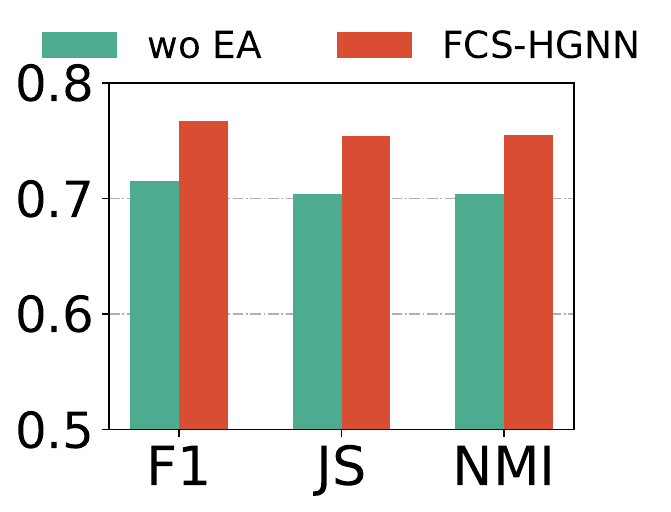} 
            \caption{ACM}
        \end{subfigure}
        
        \caption{Ablation Study of EA.}
        \label{fig:ablation_es}
    \end{minipage}%
    \begin{minipage}[c]{0.33\linewidth}
        \centering

        \begin{subfigure}[b]{0.49\linewidth}
            \centering
            \includegraphics[width=\linewidth]{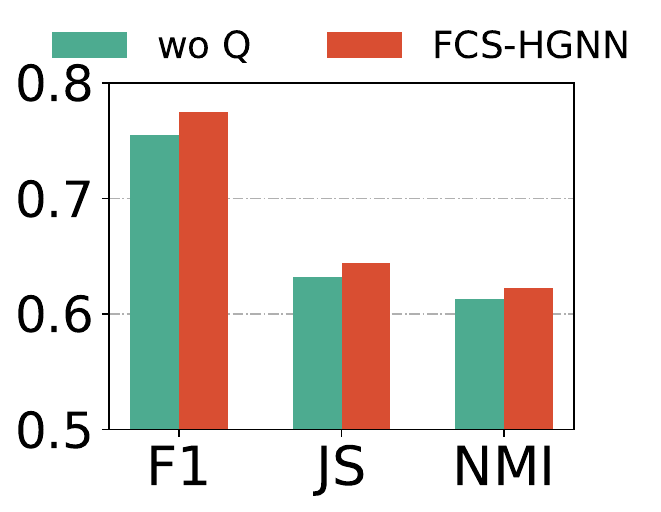} 
            \caption{DBLP}
        \end{subfigure}
        \begin{subfigure}[b]{0.49\linewidth}
            \centering
            \includegraphics[width=\linewidth]{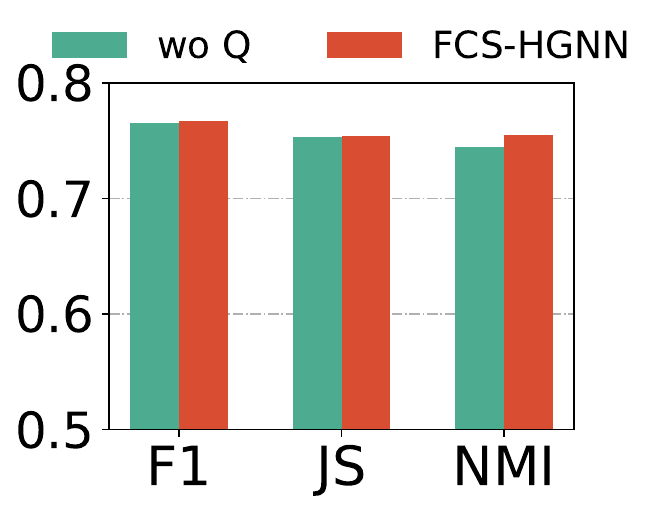} 
            \caption{ACM}
        \end{subfigure}

        \caption{Ablation Study of Query Enc.}
        \label{fig:ablation_q}
    \end{minipage}%
    \begin{minipage}[c]{0.33\linewidth}
        \centering

        \begin{subfigure}[b]{0.49\linewidth}
            \centering
            \includegraphics[width=\linewidth]{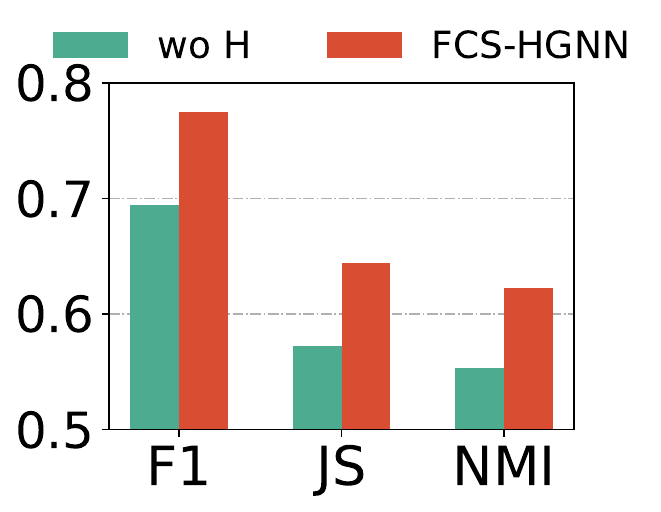} 
            \caption{DBLP}
        \end{subfigure}
        \begin{subfigure}[b]{0.49\linewidth}
            \centering
            \includegraphics[width=\linewidth]{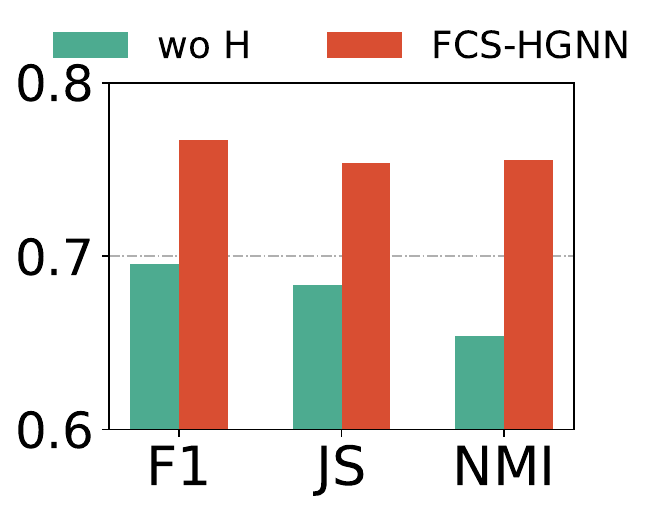} 
            \caption{ACM}
        \end{subfigure}

        \caption{Ablation Study of Hetero. Enc.}
        \label{fig:ablation_g}
    \end{minipage}%
    \vspace{-8pt}
\end{figure*}

\begin{figure*}
    \begin{minipage}[c]{0.33\linewidth}
        \centering

        \begin{subfigure}[b]{0.49\linewidth}
            \centering
            \includegraphics[width=\linewidth]{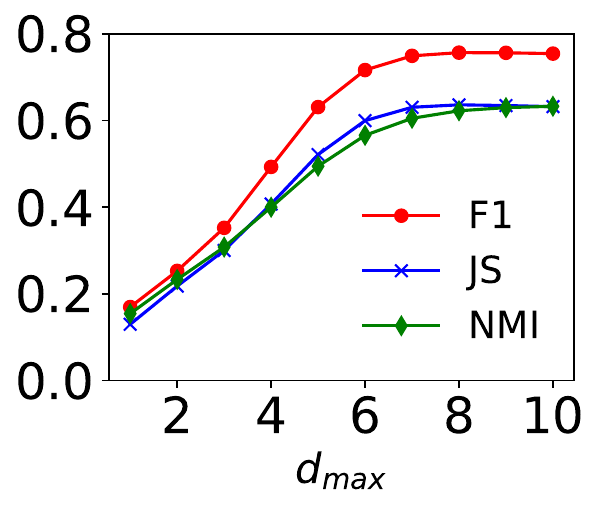} 
            \caption{DBLP}
        \end{subfigure}
        \begin{subfigure}[b]{0.49\linewidth}
            \centering
            \includegraphics[width=\linewidth]{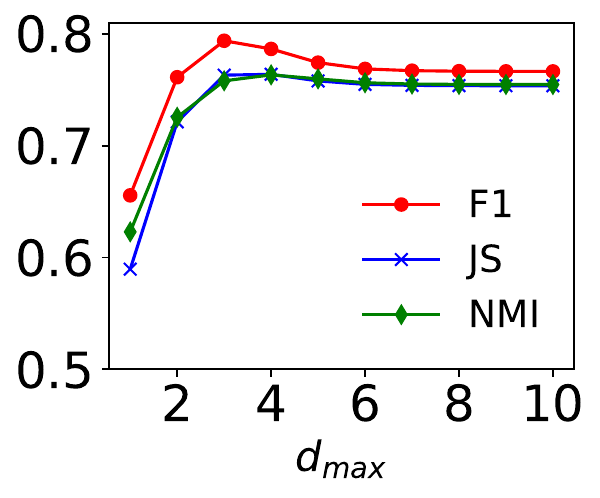} 
            \caption{ACM}
        \end{subfigure}

        \caption{Parameter Analysis of $d_{max}$.}
        \label{fig:sensitive_d_max}
    \end{minipage}%
    \begin{minipage}[c]{0.33\linewidth}
        \centering

        \begin{subfigure}[b]{0.49\linewidth}
            \centering
            \includegraphics[width=\linewidth]{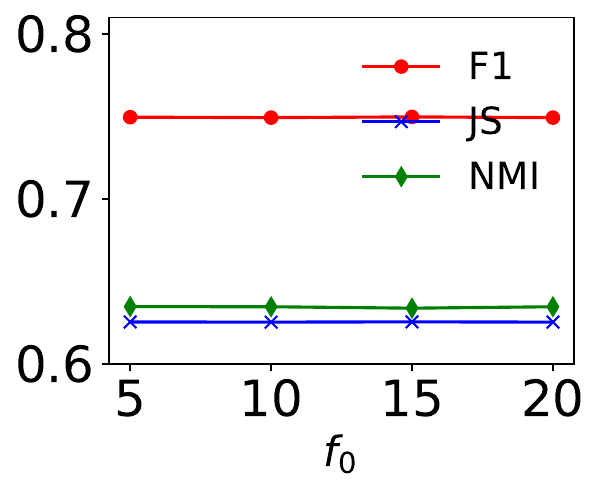} 
            \caption{DBLP}
        \end{subfigure}
        \begin{subfigure}[b]{0.49\linewidth}
            \centering
            \includegraphics[width=\linewidth]{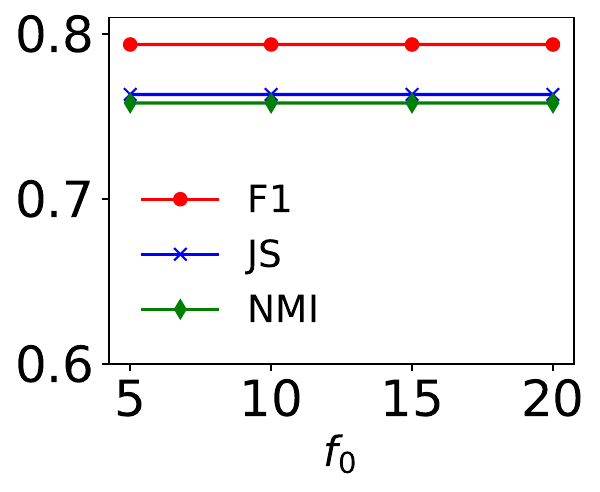} 
            \caption{ACM}
        \end{subfigure}

        \caption{Parameter Analysis of $f_{0}$.}
        \label{fig:sensitive_f0}
    \end{minipage}%
    \begin{minipage}[c]{0.33\linewidth}
        \centering

        \begin{subfigure}[b]{0.49\linewidth}
            \centering
            \includegraphics[width=\linewidth]{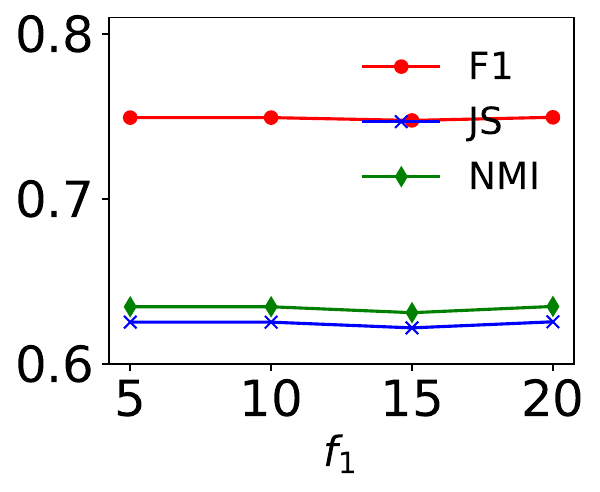} 
            \caption{DBLP}
        \end{subfigure}
        \begin{subfigure}[b]{0.49\linewidth}
            \centering
            \includegraphics[width=\linewidth]{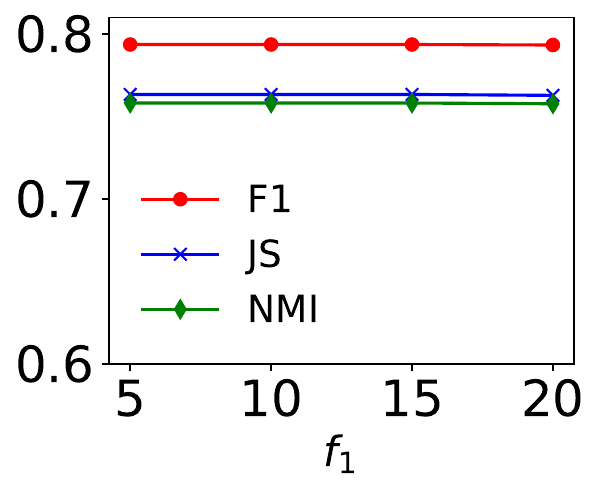} 
            \caption{ACM}
        \end{subfigure}

        \caption{Parameter Analysis of $f_{1}$.}
        \label{fig:sensitive_f1}
    \end{minipage}%
\end{figure*}

\subsubsection{Ablation Study of Edge Semantic Attention Mechanism.}
In our proposed methods, the edge semantic attention mechanism is a crucial component of both the heterogeneous encoder and query encoder. Therefore, we conduct ablation studies to evaluate the effectiveness of the edge semantic attention mechanism in community search.
We substitute the edge semantic attention with the attention mechanism mentioned in GAT~\cite{velivckovic2017graph} which only depends on nodes, and the results are shown in Figure~\ref{fig:ablation_es}, where 'wo' is an abbreviation for without, and 'EA' is an abbreviation for edge semantic attention mechanism.
It is evident that removing the edge semantic attention leads to a significant decrease in performance, even while retaining the node attention mechanism.
This suggests that edge heterogeneity is crucial for community search tasks in HINs.
Our method effectively captures and utilizes the heterogeneous information present in different edges of HINs, thereby enhancing the model's understanding of the graph structure.

\subsubsection{Ablation Study of Query and Heterogeneous Encoders.}
We investigate the query and heterogeneous encoder in FCS-HGNN to gain a deeper understanding of their individual contributions to the overall performance.
As illustrated in Figure~\ref{fig:ablation_q}, the impact of the query encoder on the accuracy of community search is relatively small, as the formation of communities primarily relies on the inherent features in HINs. However, the incorporation of the query encoder proves valuable by eliminating the need for retraining, thereby enhancing the efficiency of online queries.
As illustrated in Figure~\ref{fig:ablation_g}, the heterogeneous encoder is crucial for community search, and its absence brings significant performance degradation.
Without the heterogeneous encoder, the model relies solely on graph structure and query information to predict communities.
Since our ground truth communities are labeled by traditional community search methods~\cite{jiang2022effective,jian2020effective}, which typically rely only on graph structure, the experimental results in Figure~\ref{fig:ablation_g} still maintain relatively high accuracy.

\subsubsection{Parameter Analysis of Maximum Exploration Depth $d_{max}$.}
We investigate the key parameter $d_{max}$ in Algorithm~\ref{alg:multi_search_p}, which affects the depth of exploration.
Figure~\ref{fig:sensitive_d_max} displays the performance with varying $d_{max}$.
We observe that as $d_{max}$ increases, the metrics initially increase rapidly and then tend to stabilize.
The optimal $d_{max}$ varies with the different datasets. Overall, due to the "Six Degrees of Separation" theory~\cite{guare2016six}, $d_{max}$ around 6 can achieve the best performance. Moreover, there is not a significant performance decline as $d_{max}$ increases. Therefore, the setting of $d_{max}$ is not challenging.

\subsubsection{Parameter Analysis of fanouts $f_{0}$ and $f_1$.}
Then, we investigate the impact of the hyperparameter fanouts $f_{0}$ and $f_1$ in Algorithm~\ref{alg:neighbor_sampler}.
For the analyses, $f_{0}$ is varied within \{5,10,15,20\} with a controlled $f_1=10$, and vice versa, where $f_1$ is varied within the same range while maintaining $f_0=20$.
Figure~\ref{fig:sensitive_f0} and~\ref{fig:sensitive_f1} illustrate the performance with varying $f_0$ or $f_1$.
We observe that the model is not sensitive to the fanouts $f_0$ or $f_1$, which is consistent with our previous analysis. Due to the random sampling of neighbors in each epoch, the model is able to learn node features and community structure in an efficient and stable manner.

\section{CONCLUSIONS}\label{section:conclusion}
In this paper, we propose a novel query-driven heterogeneous graph neural network, named FCS-HGNN, for multi-type community search in HINs.
Compared to existing methods, it can flexibly identify both single-type and multi-type communities without relying on predefined meta-paths or user-specific relationship constraints, significantly alleviating the burden on users.
Moreover, FCS-HGNN dynamically consider the contribution of each relation instead of treating them equally, thus capturing more fine-grained heterogeneous information.
Furthermore, to improve efficiency on large-scale graphs, we introduce a neighbor sampling algorithm to enhance training efficiency and propose the depth-based heuristic search strategy to improve query efficiency.
We thoroughly evaluate our proposed methods on five real-world HINs and demonstrate its significant superiority in terms of effectiveness and efficiency.

\begin{acks}

This work was supported by the National Key R\&D Program of China under grant number 2023YFC3305303, the National Natural Science Foundation of China under grant number 62372434, 62302485, China Postdoctoral Science Foundation (No. 2022M713206) and CAS Special Research Assistant Program.

\end{acks}


\newpage
\bibliographystyle{ACM-Reference-Format}
\balance
\bibliography{sample-base}










\end{document}